\documentclass[11pt]{article}

\usepackage{color}
\usepackage{amssymb}
\usepackage{hyperref}
\usepackage{authblk}
\usepackage{times}
\def\prd{Physical Review D}
\def\nat{Nature}

\title{\LARGE Entropy from Carnot to Bekenstein}
\author{Ted Jacobson}
\affil{\it Maryland Center for Fundamental Physics\\
\it University of Maryland\\
\it College Park, MD 20742}

\begin{document}

\maketitle

\begin{abstract}
The second half of this essay tells the story of the genesis and early development of the notion
of black hole entropy, in the style of an after dinner talk 
(apart from a long technical footnote). The first half
sketches the development of the concept of entropy, beginning with Carnot,
passing via Thomson, Clausius, Boltzmann, Planck, von Neumann, and Shannon on the way to Bekenstein, 
and the essay ends with a retrospective on Lema\^itre. The central theme is the fault-tolerant
way that profound insights have emerged from simple (yet subtle) thermodynamic reasoning.

\end{abstract}

\section{Introduction}
Thermodynamics is a mysterious subject. Powerful conclusions can materialize out of scant inputs, 
and it is forgiving of fundamental misconceptions about the underlying details. Before recounting the tale 
of Jacob Bekenstein's discovery of black hole entropy and the generalized second law, I want to take a
few steps further back, to review the origins of thermodynamics and the concept of entropy,
and how it led to quantum physics and quantum field theory. This will help to place in perspective
the sort of process by which radical insights into new physics can come about through 
the study of thermodynamics.

\section{Heat Engines: Carnot, Thomson and Clausius}
To begin with, it is amazing that Sadi Carnot in 1824 was able to infer that there is a universal upper bound to the 
efficiency of a heat engine, even while using an incorrect concept of heat and lacking the concept of 
total energy conservation.\footnote{A brief historical account 
of the work of Carnot, and its influence on the later developments of Thomson and Clausius is given in \cite{1974PhT....27h..23K}.}
A heat engine produces a quantity of work 
when drawing heat out of a thermal reservoir and depositing heat 
into a colder thermal reservoir. Carnot reasoned that the most efficient
such engines could in principle come arbitrarily 
close to being reversible.\footnote{This requires in 
particular that the two reservoirs are not in thermal contact.}
He argued that all reversible 
heat engines, operating between the same two temperatures, 
must produce the {\it same} amount of work from the {\it same} 
amount of heat flow out of the hotter reservoir.
Were that not the case, some of the work  extracted from the more 
efficient engine could be injected to run the less efficient engine in reverse, 
while retaining the balance of the work for other purposes. 
If heat is a conserved ``caloric" fluid, then 
when the less efficient engine is run backwards, the caloric flow from 
the colder to the hotter reservoir exactly cancels the flow that occurred 
in the more efficient engine. The result, as Carnot put it,  would be 
``not only perpetual motion, but an unlimited creation of motive power without consumption
either of caloric or of any other agent,"  which he rejected as being ``inadmissable".

While Carnot's conclusion was correct, his 
argument
contained a single deep flaw: heat is {\it not} by itself conserved!
More heat flows out of the hot reservoir than flows into the 
colder reservoir, the difference being the work extracted. 
And
less heat flows into the cold reservoir with the more efficient engine than with the less efficient engine.
But now, if the less efficient engine is run backwards, the cold reservoir is no longer restored to its
initial state: more heat is drawn out than went in. The leftover work, then, is not produced from {\it nothing}, 
but rather from the heat drawn out of the {\it colder reservoir}. While not as inadmissable as 
Carnot's result, this is nevertheless inadmissable. Its impossibility is
Kelvin's version of the second law of thermodynamics. Alternatively, {\it all} of the work 
from the more efficient engine could be used to run the less efficient engine backwards, 
in which case the net result would be spontaneous (but engineered) heat flow from the colder 
reservoir to the hotter one, in violation of Clausius' version of the second law.

William Thomson (later Lord Kelvin) 
proposed in 1848 an absolute thermometric scale based on Carnot's theory, still using the 
concept of heat as a conserved caloric fluid that could not be converted to mechanical energy.\footnote{Thomson
wrote: ``The characteristic property of the scale which I now propose
is, that all degrees have the same value; that is, that a unit of
heat descending from a body A at the temperature $T^\circ$ of this scale,
to a body B at the temperature $(T-1)^\circ$, would give out the same
mechanical effect, whatever be the number $T$." 
As Klein explains: ``The temperature
$T$ on the earlier absolute scale is a
linear function of $\log \theta$, where $\theta$ is the
corresponding temperature on the later
absolute scale."}
After being convinced of the mechanical equivalence of heat by James Joule's experiments, 
Thomson in 1851 revised his proposal. 
The new version was based  on the fact that, for reversible heat engines, the ratio $Q_h/Q_l$
of the heat $Q_h$ drawn from the hotter reservoir to the heat $Q_l$ deposited in the colder reservoir 
must be not only independent of the heat engine for a given $Q_h$ (as required by the corrected
Carnot argument), but also independent of $Q_h$.  (Were the latter 
not the case, one could divide the operation into sub-processes and obtain reversible engines with different 
efficiencies.)
The ratio $Q_h/Q_l$ is therefore a property of the two reservoirs, independent of the
construction of the engine, and so it 
serves to define in an absolute fashion the ratio of the temperatures of the two reservoirs, 
$T_h/T_l:=Q_h/Q_l$. This is the absolute thermodynamic temperature
scale, which is defined up to an overall arbitrary constant multiple. 

When re-expressed as $Q_h/T_h = Q_l/T_l$, Thomson's definition of temperature 
is directly connected to  
the concept of entropy which Rudolph
Clausius developed over the decade from 1854 to 1865, when he finally gave entropy its name. 
If the entropy extracted from the hotter reservoir is defined as $Q_h/T_h$, and the entropy
added to the colder reservoir is defined as $Q_l/T_l$, then a reversible engine 
can be described as one for which there is no net change of entropy: the entropy extracted from the hotter 
reservoir is exactly equal to that deposited in the colder one. At one level, this statement is virtually empty, 
being nothing but a restatement of the {\it definition} of thermodynamic temperature in different words. 
However, because the absolute temperature can be shown to be equivalent to the ideal gas thermometer scale,
the criterion of zero entropy change has nontrivial implications for the macroscopic state variables.
Moreover, this definition of entropy change can be integrated along a sequence of equilibrium states, 
to determine the difference in entropies of the initial and final equilbrium states.
The result is independent of the sequence of states chosen to connect the two equilibria,  because 
if it were to depend on the sequence, one could contrive a means to pump heat from a colder to a hotter reservoir,
without the input of any net work, in violation of the second law. A concept of the entropy of an equilibrium state 
is thus defined, up to an overall additive constant, and up to the multiplicative constant inherent in the definition 
of absolute temperature. In a {\it non}-reversible engine,  the ratio $Q_l/Q_h$ is larger, so that more entropy 
enters the colder reservoir than leaves the hotter one, and thus entropy irreversibly increases.

\section{Atoms and Radiation: Boltzmann and Planck}

Clausius' entropy clearly had some relation to an underlying mechanical disorder, a relation to which 
Ludwig Boltzmann devoted three decades of thought, beginning at age 22 in 1866.\footnote{For 
an authoritative history of Boltzmann's work see {\it Atoms, Mechanics, and Probability. Ludwig Boltzmann's 
Statistico-Mechanical Writings. An Exegesis}, by Olivier Darrigol
(Oxford: Oxford University Press, 2018).}
 By 1871 he had obtained the expression $-\int\rho \ln\rho$ for the entropy of a dynamical system
(not necessarily in equilibrium) 
as a phase space integral, the distribution $\rho\propto \exp(-\beta H)$ being the probability density
that the Hamiltonian has the value $H$ at the temperature $\beta^{-1}$. 
The logarithm is defined only up to an additive constant, since a unit of phase space volume must be 
chosen in order to render the argument of the log dimensionless. This ambiguity was no cause for 
concern, because in any case Clausius' entropy was defined only up to an additive constant.

Boltzmann introduced in 1877 
the notion that the probability of 
a macrostate
of a gas could be identified
with the number of compatible microstates, 
equilibrium being the configuration that maximizes this number,
and entropy being its logarithm. To count the microstates he divided the 
velocity space of a molecule into small cells of uniform volume, and
described a microstate, or ``complexion", by an assignment of cells to the atoms.
He argued that the cell size is unimportant, as long as it can be chosen large enough to contain the 
velocities of a large number of atoms, while at the same time being small enough so that the
relative variation of this number from one cell to the next is small. 
The cell size then affects significantly only the overall additive constant in the entropy, which 
plays no role in thermodynamics.

At this point the stage is set for the work of Max Planck.\footnote{The motivation and evolving reasoning of Planck, as well as the reactions of his contemporaries to his work, was penetratingly analyzed by 
Thomas Kuhn in his masterpiece, {\it Blackbody Theory and the Quantum Discontinuity, 1894-1912}.
The following comments are based on Kuhn's analysis, and on the forthcoming book by 
Anthony Duncan and Michel Janssen, {\it How We Got to Quantum Mechanics}.} 
Planck set out in 1894 to establish the second law of thermodynamics as an exact statement, like the conservation of energy. He initially thought the absorption and emission of electromagnetic radiation would be a fundamentally  irreversible process that could account for an absolute formulation of the second law. At some point 
he realized this was simply inconsistent with classical electrodynamics, but nevertheless this is the concern that
led him to the blackbody problem. Despite what we are told in textbooks and colloquia, the ``ultraviolet catastrophe" was {\it not} the problem he set out to solve. (That phrase was introduced in 1911 by Paul Ehrenfest who was, by the way, a student of Boltzmann.) Following Boltzmann, Planck believed that entropy is well-defined out of equilibrium, and that it is maximized in equilibrium, which led him to the problem of understanding the blackbody spectrum from first principles. 

Planck developed over a period of five years the theory of 
oscillators (``resonators") driven by a stochastic electromagnetic field and damped by radiation reaction.
One product of this work was the equilibrium relation $\rho_\nu = 8\pi\nu^2U_\nu/c^3$, 
between the energy density per unit frequency interval 
$\rho_\nu$ of the radiation field, and the average energy $U_\nu$ of a single damped oscillator with 
frequency $\nu$. This reduced the problem to the one of finding the equilibrium distribution of $U_\nu$ values.

Wien had introduced a theoretical distribution law for $\rho_\nu$ which fit the experiments reasonably well at the time,
and in 1899 Planck set out to account for Wien's distribution using thermodynamics.
Taking Wien's law as given, Planck inferred the corresponding temperature dependence of $U_\nu$, and 
from that  he inferred that 
the entropy $S(U)$ of one resonator, as a function of its energy $U$, must satisfy the relation
$S''(U) \propto - 1/U$. He further showed that this relation is implied 
by the entropy maximization principle, Wien's displacement law, and 
the assumption that the entropy of a collection of $N$ resonators of a given 
frequency depends only on their total energy, and not on $N$. 
This was all classical physics. 

Soon came more data, however, 
extending the spectrum further into the infrared, and differing from the Wien distribution.
This led Planck to realize that his assumption of the $N$-independence of the entropy was 
unjustified, and 
he quickly found the simplest modification of his formula for $S''(U)$ that would agree with the
observed fact that, for each frequency, at sufficiently high temperature $T(=1/S'(U))$,  
the radiance is proportional to temperature. This led him directly to 
$S''(U) = - a/(U(1+bU))$, which produced a spectral density that fit all of the data very well; 
in fact it produces precisely the Planck distribution!
The equilibrium distribution of radiation was thus determined by two fundamental physical constants, 
$a$ and $b$.\footnote{The same two constants had already entered the Wien distribution, which is
after all just the high frequency limit of the Planck distribution, and
Planck had enthusiastically noted that, together with Newton's constant and the speed of light, these 
allowed the construction of a natural system of units.
Two years later in 1901, after having found his statistical derivation of the Planck distribution, 
he showed that consistency of equilibrium between radiation and ideal gas implied that these radiation constants
must also determine the Avogadro constant $N_A$ and the gas constant $R$ 
(or Boltzmann's constant $k=R/N_A$), and vice versa. 
This allowed him to express the elementary electronic charge, which is related to $N_A$ and $R$ via electrochemical measurements, in terms of the measured radiation constants, 
and he found agreement with the available estimates.}
A giant step had been taken, but Planck as yet had no derivation of the new entropy formula from first principles. 

For this he turned to Boltzmann's notion that entropy corresponds to the logarithm of the number of microstates
compatible with a macrostate, and that entropy is maximized in equilibrium.
Adapting Boltzmann's combinatorial counting of the microstates of gases, Planck set out to 
enumerate the number of ways that a given total energy can be distributed among a collection of resonators of different frequencies. The distribution of energy with respect to frequency that maximizes the total number of 
such ``complexions" would then correspond to the equilibrium distribution.
To this end it was necessary to adopt
what could be called an energy `bin size' (though Planck certainly did not refer to it this way), 
in order to discretize the counting.
To maintain agreement with Wien's displacement law, 
he took the bin size to be proportional to the natural frequency $\nu$ of the resonators, 
$\Delta E=h\nu$, which required him to introduce a proportionality factor $h$ with dimensions of action.
This quickly led to the desired result for the entropy, at the end of 1900.
As Kuhn convincingly argues,\footnote{To all the evidence discussed by Kuhn, 
and by Duncan \& Janssen, I would add Planck's Nobel Prize lecture in 1920, 
\url{https://www.nobelprize.org/prizes/physics/1918/planck/lecture/}. In that lecture, Planck nowhere mentions
having quantized the energies of the oscillators. Rather, he refers to `` `elementary regions' or `free rooms for action' ". He does
describe the difficulties fitting his derivation into classical physics, but there he is (I believe) referring to subsequent developments.} 
contrary to what we are taught in school, 
and to what his early readers like Lorentz and Ehrenfest thought he was doing, 
Planck was most certainly {\it not} assuming 
that the energy of a resonator was itself quantized. In fact, 
he did not initially think that a break with classical physics was required.

Although Planck of course recognized that the observed radiation distribution depended on the 
constant $h$ he had introduced to specify the bin sizes,
he resisted the conclusion that 
classical physics was unable to account for that formula, despite the fact that
the classical physics was complete without that constant. 
Eventually, he was convinced in 1908, by Lorentz (who had himself just come around,
and who was not the only nor the first to note), that classical electromagnetic theory
cannot produce a sensible equilibrium distribution. In the meantime, only a few 
people, beginning with Ehrenfest and Einstein, saw that a break with classical 
physics was required. And since classical equipartition of electromagnetic 
energy, or maximization of the field entropy, 
led inevitably not to the Planck distribution but to the 
Jeans distribution at all frequencies, there must have been a flaw in Planck's reasoning 
if taken classically. 
In fact, Einstein had first noted the problem in 1906: 
the Planck distribution entailed a significant
variation of the probability across one energy bin when $h\nu/kT$ is not much smaller than unity.
By the end of the decade the conclusion was widely appreciated:
only by quantizing the energy of the modes of the radiation field, and maximizing
their entropy at fixed energy, could the Planck distribution be consistently derived.
To account for a sensible theory of thermal equilibrium of electromagnetic radiation,
quantum mechanics and quantum field theory had to be discovered!

\section{Quantum Mechanics and Information Theory}
Along with quantum mechanics came a radical shift in the nature of the 
probabilities that figure in the statistical definition of entropy, but 
it wasn't until a quarter century after Planck's groundbreaking work
that this became fully clear.
Quantum mechanics demanded 
a new role for probability in physics: that of governing 
indeterminism.\footnote{In his Nobel Prize lecture 
[\url{https://www.nobelprize.org/prizes/physics/1954/born/lecture/}]
Max Born gives an overview of the emergence of the 
statistical interpretation of quantum mechanics. 
For an in depth historical look, focusing on 
the approaches of Jordan and von Neumann, 
see Ref.~\cite{2013EPJH...38..175D}.} 
Classical states were replaced 
by quantum state vectors, whose squared amplitudes
give fundamental probabilities (whatever those are). 
Moreover, because the state spaces of 
composite systems combine by 
tensor product, a generic quantum subsystem 
cannot be assigned a
definite quantum state vector.
This phenomomenon of nonseparability, or entanglement of 
bipartite quantum states,
was noted very early on
by Lev Landau in 1927 (at the ripe old age of 19), in a paper
on radiation damping of an electric dipole \cite{1927ZPhy...45..430L}.
The opening lines of the paper read:
``In wave mechanics, a system can not be uniquely defined; we always have to do with a probabilistic ensemble (statistical conception). 
If the system is coupled with another one, its behavior has a double indeterminacy." Landau then proceeds to explain algebraically what is meant.
%``In der Wellenmeehanik kann ein System nicht eindeutig definiert werden; wir
%haben es immer mit einer Wahrseheinllchkeitsgesamtheit zu tun (statistische Auffassuug)  Ist das System mit einem anderen gekoppelt, so tritt in seinem Verhalten eine doppelte Unbestimmtheit auf."
The same year, John von Neumann introduced the 
entropy $S=-{\rm Tr}\rho\ln\rho$ associated with the 
density matrix $\rho$ of a quantum system, in his paper
``Thermodynamics of Quantum Mechanical Ensembles" \cite{Neumann1927}.
Five years later, in {\it The Mathematical Foundations of Quantum Mechanics}, 
von Neumann attributed to Landau's paper  the notion that, in quantum 
mechanics, a subsystem of a pure quantum state can have entropy.

Our last station on the road to black hole entropy is information theoretic entropy, introduced by Claude Shannon in 1948 \cite{Shannon:2001:MTC:584091.584093}. For practical reasons, it was of interest to Shannon to quantify the information that can be transferred via a communication channel. The connection to thermodynamic entropy is that the microstate of a physical system can be viewed as a message whose information content is greater when the entropy of the corresponding macrostate is greater. Although quantum mechanics played no role in Shannon's considerations, 
the shift of focus from the state of the system to the state of knowledge of the communicator is in a sense required 
by quantum mechanics, at least if quantum probabilities are conceived as referring to information possessed by an observer. But the information theoretic interpretation of entropy further widens the application of this concept, by detaching it from the particulars of the thermodynamics, phase space, or Hilbert space of a specific physical system, and focusing on the information acquisition of an agent who interacts with that system. As a result, the concept of entropy can be applied even when the nature of the system is only partially known.  It is presumably for this  generality, as well as its relation to the information acquired by an observer of a quantum system, that Jacob Bekenstein in 1972 reached for the information theoretic definition of entropy when trying to determine the entropy that should be assigned to a black hole.\\

\section{Black Holes and Bekenstein}
The genesis of black hole entropy has been traced to the famous story of John Wheeler and his graduate student, Jacob, discussing teacups and black holes.
As Wheeler recalled it, the conversation took place in his office. In his 1990 book, {\it A Journey into Gravity and Spacetime}, he writes, ``I told him [Jacob] of the concern I always feel when a hot cup of tea exchanges energy with a cold cup of tea."  

I wanted to get to the bottom of this tea thing.\footnote{The following part of this article is closely 
based in part on an after-banquet speech 
given on the occasion of the conference, {\it 40 Years of Black Hole Thermodynamics}, held in honor of Jacob Bekenstein at the Institute for Advanced Studies, Jerusalem, September, 2012. Here I have maintained the 
tone and style that was appropriate to that occasion. In particular, some of the text is written in first person.} It was kind of confusing to me thinking of two cups of tea exchanging heat with each other. I mean what would be the arrangement to ensure good thermal contact? And anyway, how often do you find a cold cup of tea so close to a hot one? And why not just talk about the heat going from the tea into a single teacup when it was poured? 

I found another reference, Wheeler's 1998 book: {\it Geons, Black Holes, and Quantum Foam: A Life in Physics}, written with Ken Ford. An Amazon.com search in the book yielded 8 passages in which the word ÒteaÓ appears --- one being the index entry for the Fine Hall tea at Princeton. The first passage was about breaking for tea after discussions with Bohr, and mentioned that Bohr would lift the edge of the rug in his office and kick the chalk bits under the carpet, to avoid being scolded by the janitor. Several were about tea in Fine Hall, and discussions there with physics luminaries. Wheeler says, ``Bohr and I were regulars at the afternoon teas." He quotes Oppenheimer as saying that ``Tea is where we explain to each other what we do not understand." Another entry mentions that Einstein invited Wheeler's general relativity class to tea at his house, during the last two years of Einstein's life. So it is clear that tea played a very important role in Wheeler's scientific life\dots

And then there is the legendary discussion with Jacob. 
Wheeler writes: 
\begin{quote}
\it The idea that a black hole has no entropy troubled me, but I didnÕt see any escape from this conclusion. In a joking mood one day in my office, I remarked to Jacob Bekenstein that I always feel like a criminal when I put a cup of hot tea next to a glass of iced tea and then let the two come to a common temperature. My crime, I said to Jacob, echoes down to the end of time, for there is no way to erase or undo it. But let a black hole swim by and let me drop the hot tea and the cold tea into it. Then is not all evidence of my crime erased forever? This remark was all that Jacob needed. He took it seriously and went away to think about it.
\end{quote}
There are several notable and, frankly, puzzling things about this passage.  First, Wheeler's 1990 ``concern" has escalated into ``feeling like a criminal" because of his facilitation of an increase of entropy.  I read this passage to my wife, who is not a scientist. She asked me why did he feel like a criminal, if he was not breaking any laws? Good question. In fact, the second law of thermodynamics was fully upheld when the heat flowed from hot to cold, and the entropy increased. The crime, if there were one, would have been dropping the teacups into the black hole---{\it then} he might be breaking a law, namely the second law.
Next, notice that the cold cup of tea has now become a glass of iced tea, further complicating the situation. I mean, wouldn't the iced tea really absorb more heat from the room temperature air than from the hot tea? 
From a physics point of view, however, the most interesting thing to me is that Wheeler said that he and others thought at the time that, since a black hole has zero temperature, it must have zero entropy. I guess Wheeler was thinking a zero temperature black hole was like a zero temperature material system which, if it had a unique ground state, would have zero entropy. But, unlike a material system, the black hole could absorb energy and still remain in equilibrium at zero temperature. One should therefore have thought it would have {\it infinite} entropy, 
because any time it absorbed heat $Q$ its entropy would rise by the infinite amount $Q/0$.

Had I been there with Wheeler in his office that day, I think I would have said ``What's the problem? The entropy went down into the black hole, so what?" And I may have tried to argue that since the black hole has zero temperature, it has infinite entropy, and therefore the second law, including the black hole entropy, is upheld.
But, as Jacob appreciated, the problem is deeper than that: what good is the 
second law if you can't use it? It presumably applies to the accessible world, and nothing that goes on inside the black hole is accessible to those who remain outside.
Moreover, Jacob believed that, like all other entropy, black hole entropy must be a measure of missing information.  
If a black hole really had infinite changes of entropy, then in any process an infinite amount of information would be lost, which sounds absurd. It should be possible to lose a finite amount of information, that is, to add a finite amount of entropy. This argument suggests that a black hole therefore must have a nonzero temperature. But Jacob didn't make the argument this way. 

In an interview in the Israeli newspaper Haaretz [from June, 2012] 
Jacob spoke of his feelings about the ``laws and order of physics". He is quoted there as saying:
``I get a sense of security that not everything is random and that I can actually understand and not be surprised by things."
Thermodynamics is the order that emerges from underlying randomness. I guess that in seizing upon Wheeler's question, Jacob was on a quest to tame the randomness of the world. He insisted on the second law, even when black holes are present, somehow knowing in his heart that the law must be truly fundamental. 

The only hope for saving the second law in the presence of a black hole was to assign a finite entropy to the black hole itself. Roger Penrose had shown in 1969 that rotational energy can be extracted from a black hole \cite{1969NCimR...1..252P,2002GReGr..34.1141P}, 
and together with R.\ M.\ Floyd found in 1971 that, exploiting the fissioning of a body into two fragments, one with negative Killing energy and the other with positive Killing energy,  such energy extraction always results in an increase of the horizon area \cite{1971NPhS..229..177P}. 
In parallel, Demetrios Christodoulou had shown in 1970 that in such processes a quantity he called 
the {\it irreducible mass}---which is in fact proportional to the horizon area---cannot decrease \cite{1970PhRvL..25.1596C}.
And in 1971 Stephen Hawking proved a general theorem demonstrating that the horizon area cannot decrease \cite{1971PhRvL..26.1344H}. 

The obvious analogy of these results with the second law of thermodynamics suggested to Jacob that 
a black hole be assigned an entropy proportional to its horizon area \cite{1973PhRvD...7.2333B}. But entropy is dimensionless (when temperature is assigned units of energy), so it was necessary to divide the horizon area by a universal constant with dimensions of area, and for this only one candidate presented itself: the (tiny) squared Planck length, $\hbar G/c^3$, which is equal to about $(10^{-33}{\rm cm})^2$.  Jacob remarked that the appearance of $\hbar$ in the entropy ``is not totally unexpected", since ``the underlying states of any system are always quantum in nature", and 
he proposed the black hole entropy formula 
\begin{equation}\label{S}
S=\eta A/(\hbar G/c^3),
\end{equation}
where $\eta$ is a dimensionless proportionality constant.  As to the value of $\eta$, he wrote that
\begin{quote}
\it \dots it would be somewhat pretentious to attempt to calculate the precise value of the constant $\eta$ without a full understanding of the quantum reality which underlies a ``classical" black hole.
\end{quote}          
Instead, he provided an {\it estimate} of $\eta$, using an argument which also provided an independent
derivation of the entropy formula (\ref{S}), using quantum mechanics and the information theoretic interpretation of entropy.

Jacob reasoned this way: 
The entropy of a black hole should correspond to the information it is hiding. When one additional bit of information
is added, the entropy should increase by $\ln 2$, and there must therefore be 
a smallest nonzero amount of area that can be added {\it with certainty} to a black hole horizon, corresponding to one bit of information. In classical physics there is no lower limit. But he showed that, in view of the interplay of the gravitational redshift and the Heisenberg uncertainty relation, there is indeed in quantum physics a smallest area that can be added with certainty, and it has always the universal value $ \sim \hbar G/c^3$, regardless of the mass, spin and charge of the black hole. Having thus determined that one bit of additional entropy corresponds to one square Planck length of area, he had established (\ref{S}) from ``first principles", and deduced that $\eta$ is a number of order unity.
He could then compute, using the Clausius relation, the ``effective" temperature that should be assigned to a 
black hole. Were it not for quantum mechanics, the entropy would have been infinite and the temperature would have been zero. With quantum mechanics, the entropy was finite but huge, and the temperature was non-zero but tiny---indeed proportional to $\hbar$ (and in fact equal to the Hawking temperature
up to an unknown order unity factor).\footnote{ 
Let me briefly explain in this technical footnote the argument just described.  It
made use of the first law of black hole mechanics which, as far as I know, was first derived by Jacob. 
He found that law simply by varying the mass, angular  momentum, and charge parameters in the formula for the horizon area of a Kerr-Newman black hole. The result, with zero charge variation, can be written as 
$\delta A = (8\pi G/\kappa)(\delta M -\Omega_H\delta J)$, 
in units with $c=1$. Here $M$ is the mass, $J$ is the angular momentum, $\Omega_H$ is the angular velocity of the horizon, 
and $\kappa$ is the surface gravity. Jacob had a formula for $\kappa$ in terms of the mass, spin and charge of the black hole, but
he did not then know that this quantity is the surface gravity, which is an ``intensive variable" that is constant on the horizon.  
That was shown by Bardeen, Carter and Hawking in 1973 \cite{1973CMaPh..31..161B}, who called it the ``Zeroth law". 
To minimize $\delta A$, one should thus minimize $\delta M -\Omega_H\delta J$, which is the variation of the conserved quantity corresponding to the horizon-generating Killing vector $\chi = \partial_t + \Omega_H\partial_\phi$. I'll call this conserved quantity the
``boost energy", because on the horizon this Killing vector behaves like a Lorentz boost on a Rindler horizon in Minkowski spacetime.
Jacob thus considered  a particle of mass $m$ outside the horizon, with 4-velocity parallel to $\chi$, where its boost energy is
equal to $|\chi|m$, with $|\chi|$ the norm of $\chi$. This vanishes if the mass vanishes, or if the particle sits on the horizon, where $|\chi|=0$. But the uncertainty relation precludes using vanishing boost energy to make $\delta A$ vanish: 
If the particle is localized a distance $d$ from the horizon 
(along the hypersurface orthogonal to $\chi$), then $d\gtrsim \hbar/m$. If the mass is very small, then it is totally delocalized and cannot be known to have entered the black hole. On the other hand, if $\hbar/m$ is much smaller than the black hole, we can
use the approximation $|\chi|\approx\kappa d$, so that the conserved quantity is $(\kappa d) m \gtrsim \hbar \kappa$. 
Referring back to the first law, this implies that $\delta A\gtrsim (8\pi G/\kappa)(\hbar\kappa)= 8\pi\hbar G
=8\pi(\mbox{Planck length})^2$.
The smallest area increase is thus independent of the mass and spin of the black hole. 
The black hole entropy is therefore given by $S\sim A\ln 2 /8\pi\hbar G$, corresponding to 
$\eta \sim \ln 2/8\pi$ in (\ref{S}), and the  temperature is $T_{\rm bh}\sim\hbar\kappa/\ln 2$
(which is $2\pi/\ln 2$ times the Hawking
temperature, $T_H=\hbar \kappa/2\pi$).
Notice that the minimal boost energy that can be added with certainty to the black hole 
coincides with $\sim T_{\rm bh}$.}

With this understanding in hand, Jacob proposed the generalized second law (GSL) of thermodynamics: 
the sum of the ordinary entropy outside the black hole and the black hole entropy (\ref{S}) can never decrease.
To support this proposal, he argued that the black hole entropy should represent the inaccessible information 
about how the black hole might have formed, so that if a body with common entropy $S_{\rm c}$ is dropped into 
the black hole, $S_{\rm bh}$ must increase by at least $S_{\rm c}$, implying that the generalized entropy
must increase.  This argument makes sense provided its premise, that $S_{\rm bh}$ represents the inaccessible
information, is indeed valid. Of course he had no justification for that premise other than his deep insight and intuition
that black hole thermodynamics should make sense. But he then 
tested the GSL, without assuming that premise. The first test involved a
nonrelativistic oscillator enclosed in a box at finite temperature and dropped into a black hole, and he showed that
the generalized entropy increases. The second test involved a beam of thermal radiation aimed at the black hole.

Now the story takes a surprising twist. 
Of course the black hole absorbs radiation from the thermal beam 
and grows, increasing its entropy. 
But there is a problem:  If the beam temperature is lower than $T_{\rm bh}$, 
then heat flows from cold to hot! The GSL is violated because the beam entropy 
decreases by more than the black hole entropy increases. 

In retrospect, we can see that to save the GSL there was only one way out: to suppose that the ``effective" temperature 
$T_{\rm bh}$
of the black hole is in fact an {\it actual temperature}, so that the black hole must emit thermal radiation, and therefore entropy. 
However, Jacob did not make that argument. In his PRD paper he instead pointed out that, if the bath were colder than the black hole, the entropy-decreasing process of absorbing radiation would involve stochastic
fluctuations, because the wavelength of the thermal radiation would be much larger than the black hole.
And then came a crucial error: he asserted that in the domain of such fluctuations ``the second law is irrelevant." His deeply felt, and correct, conviction---that the generalized second law must hold---led him to dismiss a counter example using a spurious argument. 

Why did Jacob not instead infer that 
a black hole {\it must} have a real temperature, proportional to $\hbar$?
I am fascinated and puzzled by this. I'm sure he understood these matters very well. While the second law does not hold at the level of individual fluctuations, it holds on average,
as he himself stated in the very same paper.
So, why did he not predict black hole radiation in order to save the GSL?

Jacob stated in the PRD article that to assign a black hole a {\it real} temperature ``can easily lead to all sorts of paradoxes"\dots but he doesn't say what those paradoxes might be. He states in his book, {\it Buchi Neri} (apparently published only in Italian), that he wrote up the article in between lectures at the famous Les Houches summer school in 1972. He recalls a long talk he had with Brandon Carter there, writing: 
\begin{quote}
\it He stressed that since a black hole absorbs radiation perfectly, there can be no option but to assign it zero temperature, which contradicted my claim that the black hole has a real temperature. I almost lost my confidence then, and this accounts for the somewhat weak statement about black hole temperature I make in my paper\dots
\end{quote}
But while it is perhaps true that nothing can escape from a black hole, there is another, overlooked possibility:  radiation could escape from the {\it outside}, with no causality violation, as Hawking discovered in 1974 \cite{1974Natur.248...30H}. 

It seems that what was missing from Jacob's thoughts at the time is just one fundamental point: even in ``empty space," a black hole is immersed in the vacuum of quantum fields. 
I conjecture that, although he thought a lot about field quanta falling into a black hole, he never thought about the quantum field {\it vacuum} falling into a black hole. Had that thought crossed his mind, one day in 1972, I venture to guess that he would have immediately realized that the vacuum fluctuations give the black hole a {\it real} temperature, which could uphold the GSL in all settings, even when fluctuations dominate.  I'd also guess that Jacob would have gone on to conceive the information-theoretic notion that black hole entropy arises from quantum entanglement of those vacuum fluctuations on either side of the horizon, an idea only suggested ten years later by Rafael Sorkin \cite{2014arXiv1402.3589S}.
As it happened, Hawking discovered this real temperature  two years later while investigating other matters entirely. Because Hawking's calculation doesn't involve ``the quantum reality which underlies a classical black hole," but rather only the quantum reality of test fields on a
classical black hole background, it turned out that Jacob's pessimism regarding the possibility of computing the 
proportionality constant $\eta$ was unfounded. Indeed, Hawking's calculation showed that $\eta=1/4$.

\section{Cosmology and Lema\^itre}

This story of horizons and the entropy of quantum fluctuations can actually be traced back an additional forty years before Jacob's paper, to the work of  Georges Lema\^itre, a physicist, mathematician, and Roman Catholic priest from Belgium. 
Although is it not commonly known, in 1933 
Lema\^itre was the first person to understand the nature of a black hole horizon, long before it was called a black hole. He realized that a black hole horizon is locally equivalent to a cosmological de Sitter horizon \cite{1933ASSB...53...51L,Lemaitre:1933gd}. Other than Felix Klein, who had pointed to the globally nonsingular de Sitter hyperboloid, Lema\^itre was the first to understand that one can fall freely without disaster across a de Sitter horizon.\footnote{de Sitter, Hermann Weyl, and Einstein all thought de Sitter space was somehow singular at the horizon \cite{LuminetBook}.}  He showed this using a coordinate transformation, and he applied the same type of transformation to understand the black hole horizon. 
As later shown by Gary Gibbons and Hawking \cite{1977PhRvD..15.2738G}, 
de Sitter horizons have an entropy and a temperature, much as do black hole horizons.

Lema\^itre was ahead of his time in addressing seriously, 
with all the currently available physics,  
the questions raised by cosmology.
He deduced the distance-redshift relation, 
and using available data inferred a value for the constant in ``Hubble's law" for the universal expansion before Hubble did \cite{1927ASSB...47...49L,2013GReGr..45.1635L,2015arXiv150308304L,2011Natur.479..171L}, argued that without a cosmological constant the universe would be too young to accommodate the geological age of the earth (given the universal expansion rate, which at that time was incorrectly estimated to be around seven times its actual value), and formulated a theory of the origin and growth of structure. 
Noting that entropy increases according to the second law as the universe evolves, 
Lema\^itre believed that there must have been an initial state of zero entropy. 
He wrote a visionary paper about this in 1931.

Sir Arthur Eddington had mentioned this issue of zero initial entropy, in a 1931 address to the Mathematical Association, titled
``The End of the World: from the Standpoint of Mathematical Physics,"
and published in the scientific journal Nature. As the title suggests, the main subject was the infinite future and the so-called Òheat deathÓ of the universe. But he also discussed the beginning of the universe, saying: 
``We have come to an abrupt end of space-time---only we generally call it the beginning"\dots and he opined, ``Philosophically, the notion of a beginning of the present order of Nature is repugnant to me." 

Lema\^itre responded to this in a brief letter to Nature (May 9, 1931), less than half a page long. 
On the same page of the journal with a letter from British Columbia on the chemistry of cheese ripening, 
and another letter with a photograph of insects found in the gut of a cobra in Malaysia, 
is Lema\^itre's letter, titled:
``The Beginning of the World from the Point of View of Quantum Theory".
In the letter he traces the universe back to the ground state of a primeval atom. He writes 
\begin{quote}
\it \dots it may be that an atomic nucleus must be counted as a unique quantum, 
the atomic number acting as a kind of quantum number.
If the future development of quantum theory happens to turn in that direction,
we could conceive the beginning of the universe in the form of a unique atom, 
the atomic weight of which is the total mass of the universe. 
This highly unstable atom would divide in smaller and smaller atoms by a kind 
of super-radioactive process.
\end{quote}
Lema\^itre goes on to suggest that ``Some remnant of this process might, according to Sir James Jeans's idea, foster the heat of the stars until our low atomic number atoms allowed life to be possible." We now know that in fact the elements evolved in the {\it opposite} direction: the actual source of stellar energy is not radioactive fission, but rather the fusion of light elements into heavier ones.  And life could evolve only once stars that forged these elements exploded in supernovae, scattering their nuclear progeny for future chemistry. But the notion that the universe we see originated from
the decay of a symmetrical, zero entropy quantum state is perfectly in line with our best current understanding.

Looking back at the beginning of time, Lema\^itre squarely faces what could be called the cosmic information problem: the initial, pure quantum state of his primordial atom would be unique, with no entropy. Quantum mechanics would preserve this purity, so from whence could the variety of all creation be found? Lema\^itre had no hesitation answering this question. He wrote:
\begin{quote}
\it Clearly the initial quantum could not conceal in itself the whole course of evolution; but, according to the principle of indeterminacy, that is not necessary. [\dots] the whole story of the world need not have been written down in the first quantum like a song on the disc of a phonograph. The whole matter of the world must have been present at the beginning, but the story it has to tell may be written step by step.
\end{quote}
To my eye, this 
bears an
uncanny resemblance to the modern account of primordial cosmology. The primeval atom of Lema\^itre's intuition can be identified with the unstable vacuum of inflationary cosmology. The early universe, according to inflation, was a symmetrical, pure quantum state, the de Sitter vacuum, whose vacuum energy spontaneously decayed into a plasma of matter and radiation. 
The structure we see today emerged from that zero entropy vacuum as a result of quantum indeterminism.
It arose via gravitational instability, seeded by tiny deviations from equilibrium which had descended  
from nothing but primordial vacuum fluctuations. 

\subsection*{Acknowledgments}

I am grateful to  B.~Banihashemi, S.~De Baerdemaeker, J.~R.~Dorfman,  J.~Uffink, and especially M.~Janssen for 
comments, suggestions, and help with historical accuracy. This work was supported in part by NSF grant PHY-1708139.

\bibliographystyle{unsrt}
\bibliography{CarnotBekenstein}

\end{document}